\documentclass[11pt]{article}
\usepackage{geometry}
\usepackage{graphicx}

\title{16 Propositions to Reconsider\\ the Organization of a Scientific Workshop}
\author{by Christoph Schommer, University of Luxembourg}

\date{\today}

\begin{document}
\maketitle
\tableofcontents


\section{Motivation}
Participating a scientific workshop is nowadays often an adventure because the number of participants do seldom exceed the number of talks. A half-day workshop is mostly finished at lunchtime, speakers are sometimes not present and unexcused, and a strict progression of the workshop offers little air for discussion. And when talks are re-scheduled on short notice in case that a speech is dropped out, attaining guests definitely wonder why the presenter is talking about something that does not match the previously announced talk.

In our opinion, the idea of a basic understanding of a workshop has changed; it is less a meeting place of researchers who share a common research interest but more a market place where scientific papers are to be submitted and published - for a fee. It is still a major goal to increase publications, to either promote his own career or to travel around the world. A communicating interest and an active participation is often annoying. Depending on the culture and on career management and planing, the number of publications is a consequential facet, resembling the collection of decals. Moreover, the workshop does not support the different expert levels that exist in the targeted research field but is - in this respect - aimed undirected. As for example, it is commonly accepted that young researchers - being shy, often unexperienced and new in business - can not really follow presentations in a sufficient way.

Therefore, we believe that the organization of a workshop in the classical sense must be reconsidered. It is not enough of compelling the presenters to pay the registration fee only and to let the participants being impassive or taken away mentally. New concepts are needed so that this article is with some ideas to support not only the establishment of workshops being interesting and a participation more worthwhile, but rather to increase the value of an exchange of interests. In this respect, we therefore address some ideas to become implemented in the future workshop organization in order to identify workshops as a place of interaction.

\section{Propositions}
The following concepts are partitioned into three chronological steps: concepts that concern with the time before the workshop, with the time during the workshop, and with the time after the workshop. We refer to author as the persons who have submitted the paper directly and to co-authors who additionally appear on the document. The presenter is the person who is identified as to present the paper contribution. 
 
\subsection{Communicating the Motivation}\label{motivation}
During the submission of a scientific paper contribution, each author must give a short motivation why the paper has been submitted to this workshop. This must be part of the submission procedure and be taken into account for the evaluation. As it is commonly handled, at least one author must pay the registration fee and must declare to be present at the workshop. Alternatively, the authors may organize someone else who can take this over or agrees to use new media (see \ref{media}). If the presenter fails unexcused, all authors and co-authors are blacklisted (see \ref{blacklist}). With this, the workshop can become more credible; participants may be sure that the presentation takes place.

\subsection{Reviewing}\label{reviewing}
The review process must be simplified and not been complicated by too many different questions. A solid review frame may contain a general statement of the reviewer, the final decision if the contribution is to be accepted or rejected, as well as the own confidence to the subject. 

\subsection{Establishing a Blacklist}\label{blacklist}
The information, who is intended to present the accepted scientific paper, must be given when the paper is submitted. Each author who is missing unexcused is to be blacklisted. In this context, the word unexcused means for example that the author has not announced his absence and is not prepared to make it up believable for a certain period after the workshop. All of the authors receive black points that range from 1 to \emph{k} with a maximal limit of \emph{max}. The contact persons of the accepted paper receive up to \emph{k} black points, whereas co-authors receive only 1. In case that a co-author was intended to come but did not, this person is blacklisted most with up to \emph{k} points whereas the contact persons receive only 1. If an author excesses the given limit of \emph{max} black points, he is barred from presentation for a period of time. The black points may become deleted continuously in case the author has not received black points, for example \emph{m} black points per year. In fact, the blacklist and the black points can either be arranged for a workshop by the workshop organizers or a whole conference by the conference organizers or - if an international committee can be established - by a worldwide list. The black list should not be public but each researcher must have an access to his/her own status. The major aim is to satisfy the communication between researchers, and it is absolutely unacceptable if research opinions and achievements remain uncommented.

\subsection{Re-thinking the placement}\label{organisation}
Workshops are important as they concern with current research issues much more detailed than conferences do. Often, they are associated with or integrated in conferences and in almost each cases placed marginally on the first and/or on the last day. It is comprehending that this is very disastrous: why do more general talks call the tunes and not the workshops? We therefore demand for a general comprehension and in this respect a re-organsiation of conferences: if they are combined with workshops, then workshops must be centralised, but conference talks be put on the first and last day. Whis this, invited talks become more appropriate to open and close the conference event.

\subsection{Strengthening the Lead Time}\label{dyeing}
Workshops consist exclusively of presentations that are ordered. This is a lack as some contributions live from practical demonstrations or build on other presentations. Additionally, presentations are pressed into a time schedule with a limited amount of time for questions (see \ref{timequota}). Generally, workshops start with a welcome of the organizers, which is then followed by the first presentations: however, why is there no time reserved for an initial reading or at least an extended overview of the presentations that follow? With such a preparation, it is much more easier to structure the contributions and to organize questions (even with cross-links to previous and coming presentation within the workshop). 

\subsection{Being a Godfather}\label{godfather}
Each presenter is not only a presenter of his own contribution but also a \emph{godfather} of another paper in this workshop. This certainly includes the moderation of that paper, the leading of an ongoing discussions, and the support of writing a paper summary that includes the major discussion points (see \ref{summary}). This role is to be seen different to the role of a chairman, because the moderator must include the participants' opinion as well. The major advantage here is that there is a communication established. Which contribution is moderated by which participant must be done before the workshop takes place since each godfather must concern with the contents, respectively. If a godfather is not be around during the presentation, (s)he will receive black points (see \ref{blacklist}).

\subsection{Providing Green, Yellow, and Red Cards}\label{green}
Each participant could provide a \emph{Green Card}, which is a virtual card that allows to stop the presentation for a shorter time, for example to intensively comprehend the knowledge that has been communicated or to ask for a more detailed explanation. However, the number of breaks must be limited to a small number, otherwise the presentation will not come to an end. A \emph{Yellow Card} could be an advice or indication that something is not correct with the presentation. For example, the presenter stands with his back to the audience or mumbles in a way that makes him hard to understand or an presented issue is obviously wrong. A \emph{Red Card} may automatically denote the end of the talk, for example in case that the presenter has exceeded his time limit.

\subsection{Involving Alternative Media}\label{media}
Having telecommunication invigorated with intelligent solutions in communication, a personal presence is strongly desirable. However, it might be advisable to use alternative communication channels instead, for example a video conferencing through Skype. Depending on the speed, this might be a possibility to overcome unexcused missing (see \ref{blacklist}).

\subsection{Pinging Experts hit-or-miss}\label{ping}
During the workshops, non-present experts who contribute to the clarification of circumstances could be contacted, for example by telephone, skype, or other video-conferencing machines. Such a short way surely contributes to an enrichment of the workshop. However, the expert must be informed quite early, not being surprised or away.

\subsection{Offering Time Quota}\label{timequota}
Each presenter may book time for his presentation (including the discussion) which must lie between a minimum and a maximum period. A booking beforehand has the advantage that the meeting can be planned much better.

\subsection{Summarizing the Talk and the Discussion}\label{summary}
The result of each workshop is to be summarized by some lines about the contents of the talk (regularly, it will be the abstract including keywords) but plus the major contribution, criticism, and/or ideas that came out of the discussion. This individual paper summary must be done by the godfather in cooperation with the author(s). The purpose is that this strengthens not necessarily the contact making and the scientific exchange but that it supports at least a more fruitful workshop life. As a secondary result, the summary may be used for future call for papers. In this respect, the end of the workshop is not a last presentation but the writing of each presentation summary and the establishment of a \emph{final version}.

\subsection{Keeping the Proceedings as a Gift}\label{proceedings}
Proceedings must be done after the workshop has taken place. First, the written summaries (see \ref{summary}) can taken into account. Second, contributions that were not presented by authors can be kicked out from publication.

\subsection{Pursuing Common Activities}\label{activities}
The word \emph{workshop} originates from \emph{to work out} and remembers in form and content a study group. This study group is together some time, therefore social things like lunch (should be served in situ in order to avoid to hazard the progressing of the workshop) or a walk (a common excursion) should be done together.

\subsection{Administering a Digital Library}\label{library}
To avoid that presented information and even the knowledge transfer gets lost, the written summary (see \ref{summary}) might become stored in a digital library with a public access. 

\subsection{Strengthen the Discussions}\label{discussions}
Our proposal goes much further as we claim for setting the focus more to discussions than to the presentation itself, for example at least in the same proportion. 

\subsection{Linking with a Panel Session}\label{panel}
A workshop should be joined with a panel sessions that is open to everyone without any fee. Interested participants may have the chance to direct questions that are related to the presentation or even to the research field in general. The advantage is that researchers may less abdicate from their responsibility, the visibility of own research becomes more transparent, and that the opportunities of getting interdisciplinary contacts may increase.

\section{Conclusions}
The given suggestions may support a more interactive get-together during the scientific workshop. We believe that the advancement of social factors feeds the ground for a more fruitful collaboration. Probably, the costs of a workshop increase because additional technical equipment must be prepared or software installed, but the efforts to organize and arrange a scientific exchange must be more worth than an amusement.

To motivate a collegiate participation is not a minor motivation, because students stand for the future of computer science. Being a mirror of their generation, they share different opinion about newest techniques and modern life in respect to approved and reputable discussions. To enthuse young researchers will surely be one of the most important challenges.

Of course, not all communicated hypotheses and/or scientific contributions are worth to read, but the discussion of enervating and provoking hypotheses is something that must be continuously focused on. If we still follow a publication mainstream, we will not advance. If we still be in our knowledge hull, we will never see the world outside of our field of vision.

\end{document}